%% file: main.tex
\documentclass[conference]{IEEEtran}
\IEEEoverridecommandlockouts
\usepackage{cite}
\usepackage{amsmath,amssymb,amsfonts}
\usepackage{algorithmic}
\usepackage{textcomp}
\usepackage{xcolor}
\usepackage{listings}
\usepackage{hyperref}
\usepackage{comment}
\usepackage{todonotes}
\usepackage{graphicx}
\usepackage{subcaption}
\usepackage[inkscapelatex=false]{svg}

\newcommand{\alessandro}[1]{\todo[inline]{Alessandro: #1}}

\definecolor{background}{rgb}{0.95, 0.95, 0.95} 
\definecolor{keywords}{rgb}{0.0, 0.0, 0.8}      
\definecolor{comments}{rgb}{0.3, 0.6, 0.3}      
\definecolor{strings}{rgb}{0.8, 0.1, 0.1}       
\definecolor{functions}{rgb}{0.5, 0.0, 0.5}     
\definecolor{numbers}{rgb}{0.6, 0.3, 0.0}       

\def\BibTeX{{\rm B\kern-.05em{\sc i\kern-.025em b}\kern-.08em
    T\kern-.1667em\lower.7ex\hbox{E}\kern-.125emX}}
\begin{document}

\title{Breaking Down Quantum Compilation:\\Profiling and Identifying Costly Passes}

\author{\IEEEauthorblockN{Felix Zilk, Alessandro Tundo, Vincenzo De Maio, Ivona Brandic}
\IEEEauthorblockA{
HPC Research Group, Faculty of Informatics \\
TU Wien, Vienna, Austria \\
\{felix.zilk,alessandro.tundo,vincenzo.maio,ivona.brandic\}@tuwien.ac.at}
}

\maketitle
\begin{abstract}
With the increasing capabilities of quantum systems, the efficient, practical execution of quantum programs is becoming more critical. Each execution includes compilation time, which accounts for substantial overhead of the overall program runtime. To address this challenge, proposals that leverage precompilation techniques have emerged, whereby entire circuits or select components are precompiled to mitigate the compilation time spent during execution. 
Considering the impact of compilation time on quantum program execution, identifying the contribution of each individual compilation task to the execution time is necessary in directing the community's research efforts towards the development of an efficient compilation and execution pipeline. In this work, we perform a preliminary analysis of the quantum circuit compilation process in Qiskit, examining the cumulative runtime of each individual compilation task and identifying the tasks that most strongly impact the overall compilation time. Our results indicate that, as the desired level of optimization increases, circuit optimization and gate synthesis passes become the dominant tasks in compiling a Quantum Fourier Transform, with individual passes consuming up to 87\% of the total compilation time. Mapping passes require the most compilation time for a GHZ state preparation circuit, accounting for over 99\% of total compilation time.

\end{abstract}


\input{sections/01-intro}

\input{sections/02-background}
\input{sections/03-related-work}
\input{sections/04-approach}
\input{sections/05-results}
\input{sections/06-conclusions}

\section*{Acknowledgments}

F.Z. acknowledges funding by the Internet Stiftung through the Netidee scholarship ID 7413 (Optimizing Hybrid Workflows for Cloud-Based Quantum Computation). Furthermore, this research was funded in whole or in part by the Austrian Science Fund (FWF) [10.55776/PAT1668223, 10.55776/P36870] and by the Austrian Research Promotion Agency (FFG) Flagship Project HPQC (High Performance Integrated Quantum Computing) \#897481. 

We acknowledge the use of IBM Quantum Credits for this work. The views expressed are those of the authors and do not reflect the official policy or position of IBM or the IBM Quantum team. In this paper, we used ibm\textunderscore brisbane, which is of the IBM Quantum Eagle r3 processor type.

\bibliographystyle{IEEEtran} 
\bibliography{references} 

\end{document}

%% file: sections/01-intro.tex
 
\section{Introduction}\label{sec:introduction}

Quantum computing (QC) has gained particular interest in both academia and industry due to its promise to significantly speed up certain computational tasks~\cite{Nielsen_Chuang_2010, Bayerstadler2021}. Potential applications include the simulation of physical systems in materials science and chemistry~\cite{Bauer2020}, optimization problems~\cite{Abbas2024}, and machine learning~\cite{Biamonte2017}. Over the last decade, QC has become particularly popular due to significant advances that brought early prototype in-lab demonstrations~\cite{Gulde2003, DiCarlo2009} to production-grade systems available as cloud-based services~\cite{Maring2024, Karalekas2020} or integrated into high-performance computing (HPC) facilities~\cite{Beck2024, Ruefenacht}. To date, researchers have used these state-of-the-art quantum systems to run computational workloads involving more than 100 qubits~\cite{Pelofske2024, Farrell2024} and demonstrate computational advantage for specific problems using different physical platforms~\cite{Madsen2022, Wu2021}. 

Dedicated software development kits (SDKs), such as Qiskit~\cite{javadiabhari2024quantumcomputingqiskit}, are used for developing quantum algorithms and programming quantum devices. Their current workflow for the development and execution of quantum programs includes several steps, including high-level optimization, compilation, execution on hardware, and post-processing tasks~\cite{Quetschlich2025MQTPred, Leymann2020}. The compilation step, that is, the transformation of abstract quantum programs into instructions that can be executed on a quantum computer, involves numerous computationally expensive tasks, such as mapping logical qubits from an abstract circuit definition to a physical implementation on the quantum device and converting high-level circuit operations into native hardware instructions~\cite{Quetschlich2025MQTPred, Leymann2020}. By default, quantum programs are entirely compiled at each execution. Consequently, the time spent on compilation has a significant contribution to the overall runtime of the whole program~\cite{Karalekas2020, MQTPreCompQCE}, which leads to a significant runtime overhead, especially as both the size and complexity of the original circuit scale~\cite{MQTPreCompQCE}.

Recently, a number of solutions have been proposed to address this challenge through the use of precompilation methods, including~\cite{QrispQaching, MQTPreCompQCE, Karalekas2020, AccQOC, Gokhale2019, Kudrow2013}, which involve the compilation of a portion of the source code to be executed prior to deployment and stored in advance, rather than being compiled at runtime. These approaches concentrate on precompiling specific gates~\cite{Kudrow2013}, the logical circuit level~\cite{MQTPreCompQCE}, or the pulse level~\cite{Gokhale2019, AccQOC}. However, the question of which individual tasks, so-called passes, and which parameters (e.g., circuit structure, optimization level, etc.) contribute to compilation time — and to what extent — has not been addressed.


In this work, we perform a preliminary analysis of Qiskit's built-in compiler toolchain to identify which of its passes affect the compilation time for a given circuit with varying optimization levels. Consequently, we aim to help researchers and developers identify potential bottlenecks in the compilation process and to effectively use precompilation techniques. In particular, we comprehensively profile Qiskit's preset compiler pipelines, focusing on the cumulative CPU time required for individual passes and identifying the top 10 most expensive passes for a Quantum Fourier Transform (QFT)~\cite{Pattanayak2021} and a Greenberger-Horne-Zeilinger (GHZ) state~\cite{Greenberger1989} preparation circuit with 100 qubits each. Our results show that while the contributions of synthesis passes among the top 10 reveal a comparable trend for both circuits and all optimization levels, especially for higher optimization levels, the impact of qubit mapping and circuit optimization passes varies significantly between the two circuits. For instance, a single mapping pass accounts for over 99\% of the total compilation time when compiling GHZ with optimization levels 2 and 3. Similarly, for optimization level 3 and QFT, a single circuit optimization pass accounts for $\approx$87\% of the overall compilation time.

%% file: sections/02-background.tex
\section{Background}\label{sec:background}

\subsection{Quantum Program Execution Model}

\begin{figure}[!ht]
    \centering
    \includegraphics[width=0.92\linewidth]{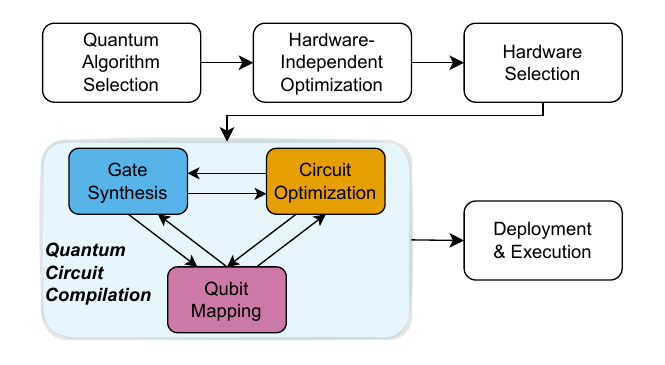}
    \caption{The depicted workflow for quantum program execution is based on the model in~\cite{Leymann2020} and adapted with a more nuanced view of quantum circuit compilation, as presented in~\cite{Quetschlich2025MQTPred}. 
    }
    \label{fig:workflow}
\end{figure}

The execution of a quantum program on a quantum computer can be modeled in a workflow consisting of several steps, as outlined in~\cite{Leymann2020} and illustrated in Fig.~\ref{fig:workflow}. First, a suitable algorithm is selected to solve the problem at hand. Next, a hardware-independent circuit optimization step is applied, followed by a hardware selection step~\cite{Quetschlich2025MQTPred}. Once a device has been selected and the physical constraints of the targeted quantum processing unit (QPU) are known, device-specific quantum circuit compilation can proceed~\cite{Quetschlich2025MQTPred}. In this phase, the circuit is compiled and optimized for the selected hardware platform on which it is to be executed. Finally, the compiled circuit is sent to a quantum computing device for deployment and execution, which returns the result~\cite{Leymann2020}.

Quantum circuit compilation (QCC) — i.e., the transformation of abstract, hardware-agnostic quantum circuits into a version optimized for the target hardware — is a non-trivial procedure involving numerous time- and resource-intensive tasks. Specifically, the abstract quantum circuit must be translated into a sequence of instructions that can be executed natively on the target QPU; a task known as gate synthesis~\cite{Quetschlich2025MQTPred}. In addition, the hardware-agnostic qubit layout of the abstract quantum circuit has to match the specific topology of the physical device; a process termed qubit mapping, a known NP-complete problem~\cite{Botea2021}. Both processes change the composition of the circuit, leading to further possibilities for circuit optimization (see Fig.~\ref{fig:workflow}). 

\subsection{Quantum Circuit Compilation in Qiskit}
Qiskit performs QCC by executing a sequence of compilation tasks, referred to as passes. These passes are typically executed through a \texttt{PassManager}, which arranges and performs a series of circuit inspections and transformations~\cite{IBMQuantumDocumentationTransp}. Specifically, Qiskit offers four built-in compiler pipelines, each designed to cater to distinct levels of optimization: 0 (no optimization), 1 (light optimization), 2 (medium optimization), and 3 (high optimization). Each of these built-in compiler pipelines is organized into multiple stages: initialization, layout, routing, translation, optimization, and scheduling. 

These stages correspond to the steps shown in Figure~\ref{fig:workflow}, in particular, hardware-independent optimization (\textit{initialization}), qubit mapping (\textit{layout} and \textit{routing}), gate synthesis (\textit{translation}), and circuit optimization (\textit{optimization}). 

The pipelines are generated by the \texttt{generate\_preset\_passmanager} function, which creates the respective \texttt{PassManager}. According to the Qiskit documentation, the standard usage of this method requires at least a \texttt{backend} instance and an \texttt{optimization\textunderscore level} as inputs to generate the pipeline; hence, we will focus on these two variables in our preliminary analysis. The key method that executes the logic of each compiler pass is the \texttt{run(dag)} method, which is implemented by each individual pass. The execution of the entire chain of compiler passes is implemented by the pass manager's \texttt{run(circuit)} method, which applies all involved passes to the quantum circuit and returns a compiled circuit that is executable on the target hardware.

%% file: sections/03-related-work.tex

\section{Related Work}\label{sec:related-work}
Existing work on quantum compilation~\cite{Kudrow2013, Gokhale2019, AccQOC, Karalekas2020, MQTPreCompQCE} concentrates primarily on the precompilation of either entire logical circuits or parts thereof with the objective of reducing compilation time during execution later on. Kudrow et al.~\cite{Kudrow2013} propose a method that aims to reduce the compilation time for arbitrary rotation gates by precompiling a specified set of rotations. In contrast, Gokhale et al.~\cite{Gokhale2019} and Cheng et al.~\cite{AccQOC} employ precompilation to mitigate compilation overhead for approaches that compile abstract circuits directly to the pulse level. Karelakas et al.~\cite{Karalekas2020} have addressed the runtime bottleneck that arises from QCC in quantum-classical cloud architectures, focusing on hybrid variational algorithms. Their proposal includes a compilation method that precompiles an abstract circuit with placeholders for gate parameters, bypassing the need to recompile the entire circuit at each iteration. Finally, the work of Quetschlich et al.~\cite{MQTPreCompQCE} has proposed a compilation approach that precompiles a generic quantum circuit that represents an entire class of problems. Their approach adapts the compiled circuit to the specific problem at runtime by solely subtracting unnecessary gates. However, while these approaches aim to reduce the required compilation time during quantum program execution, they do not consider the contribution of individual compiler passes.

Recently, the Qrisp~\cite{seidel2024qrispframeworkcompilablehighlevel} framework proposed caching for individual functions~\cite{QrispQaching} of a quantum program. Here, the Python interpreter traces a decorated function only once, ensuring that subsequent calls to the same function execute without interpreter-induced delay. However, there is currently no guideline available to determine which functions are more or less suitable for this approach.

These contributions underscore the significance of the required compilation time for quantum programs. Nevertheless, the existing literature is missing an examination of which individual passes impact the overall compilation time of a given circuit, taking into account their dependencies on parameters such as circuit structure and desired degree of optimization. 

%% file: sections/04-approach.tex

\section{Profiling Methodology}\label{sec:approach}

\begin{figure}[!ht]
    \centering
    \includegraphics[width=0.95\linewidth]{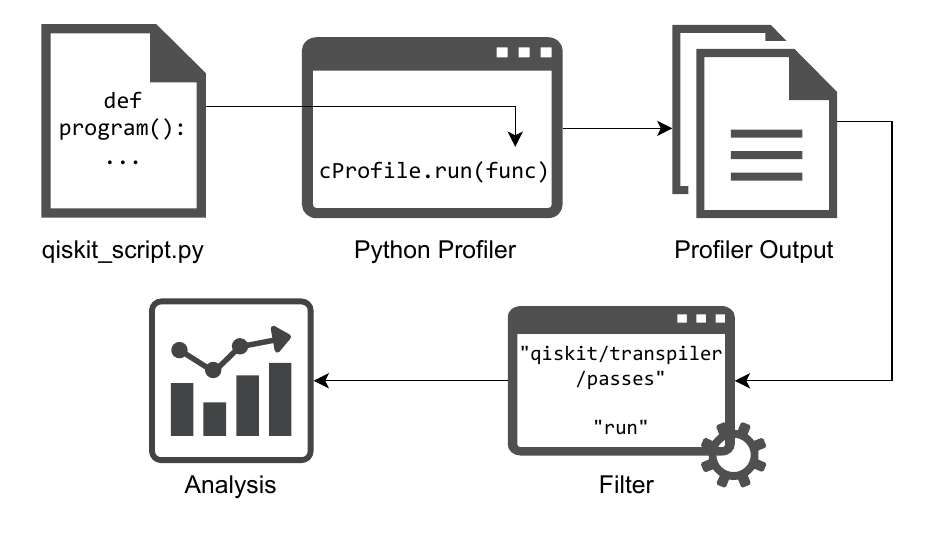}
    \caption{The workflow of our profiling methodology.}
    \label{fig:method}
\end{figure}

Figure~\ref{fig:method} illustrates our profiling methodology to monitor all compiler passes involved in a Qiskit program execution and collect their execution profile\footnote{A profile is a set of statistics that describe how often and for how long different parts of the program are executed~\cite{cProfile}.}. We choose Qiskit due to its wide adoption as an SDK for quantum programs~\cite{unitaryfund}, and both its flexibility and performance, as reported in a recent benchmark study~\cite{nation2025benchmarkingperformancequantumcomputing}. Specifically, we conduct our analysis using Qiskit SDK v1.3.2 and its preset compilation pipelines.   



We monitor the program execution using the Python profiler cProfile~\cite{cProfile}. The profiler output provides a set of statistics, including the cumulative time spent in a function and calls to all its sub-functions. From this output, we filter only for those functions that contain the target location where Qiskit's built-in passes are implemented, i.e., ``\texttt{qiskit\textbackslash transpiler\textbackslash passes}'', in their respective file path and the keyword ``\texttt{run}'' for the respective function name. Then, we extract the class names for each pass and the cumulative time spent in its \texttt{run} method and determine the 10 most costly passes with regard to overall runtime. 

We categorize passes according to the module they are implemented in, e.g., passes in ``\texttt{\ldots \textbackslash passes\textbackslash synthesis}'' as synthesis passes. To categorize passes in generic modules like \texttt{basis} or \texttt{utils}, we consider their stage of occurrence. For instance, \textit{MinimumPoint}, which is from the \texttt{utils} module, occurs in the optimization stage; thus, we assign it to the circuit optimization category. We leave passes from generic modules that occur in multiple stages uncategorized.

Our preliminary analysis includes two quantum algorithms: the QFT~\cite{Pattanayak2021} and the GHZ state~\cite{Greenberger1989} preparation. Both algorithms are implemented with 100 qubits to make sure that our test scenarios are representative of current workloads. We conduct all experiments using QPUs from the IBM platform to compare compilation times against actual quantum resource consumption. Specifically, we use the ibm\textunderscore brisbane QPU, which is an Eagle r3 processor type. We perform a total of 30 executions per configuration (per circuit and optimization level). We executed quantum programs using Qiskit v1.3.2 and Python 3.9.2, and all stages of the Qiskit compiler pipeline are executed on an HPC node with a total number of 48 available CPU cores (Intel(R) Xeon(R) CPU E5-2650 v4 $@$ 2.20GHz) with 250 GB of memory running Debian Linux 11. 

%% file: sections/05-results.tex

\section{Results \& Discussion}\label{sec:results}

\begin{figure}[!ht]
    \centering
    \includegraphics[width=0.95\linewidth]{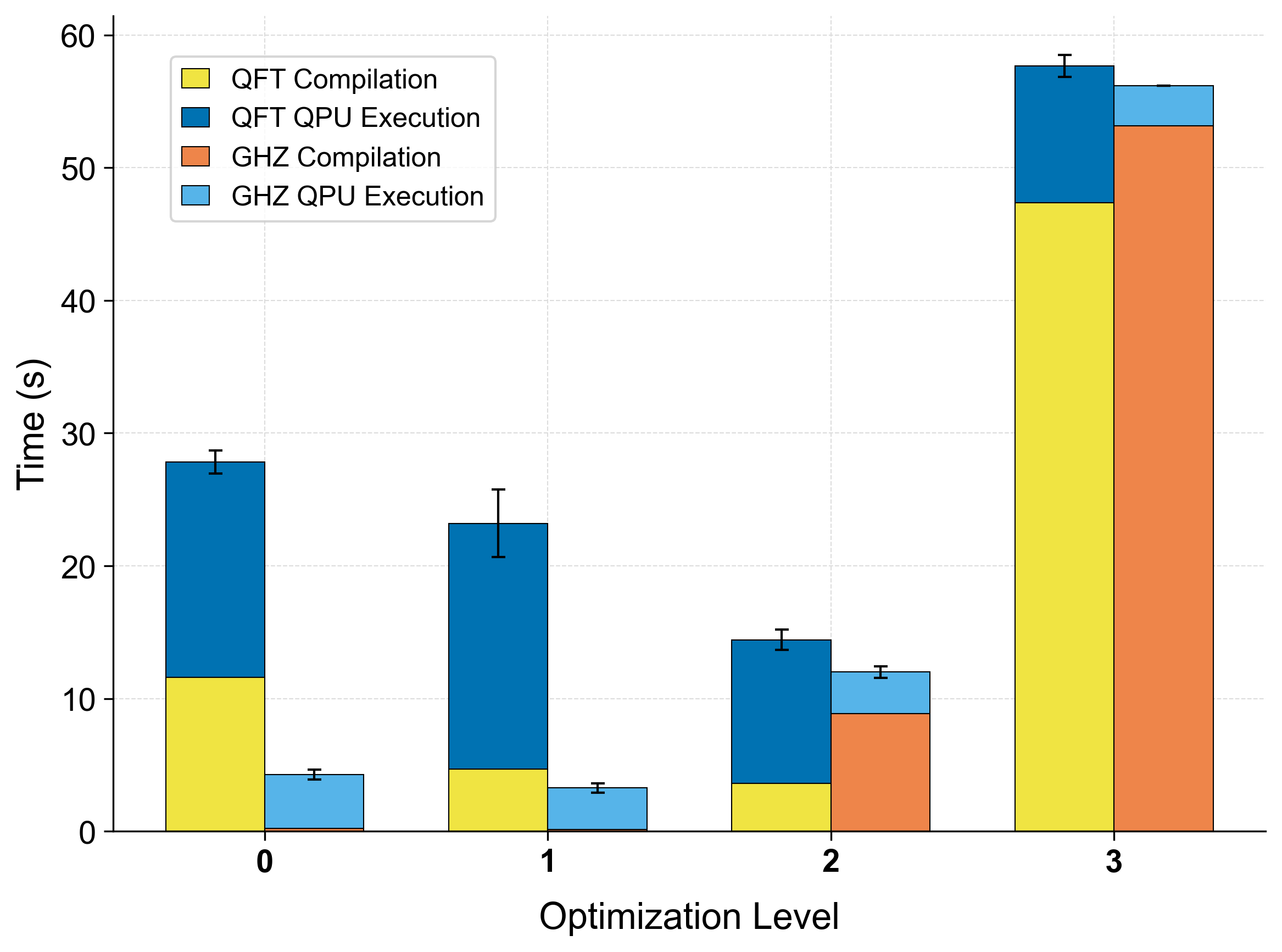}
    \caption{Compilation and QPU execution times of the QFT and GHZ circuits for different optimization levels.}
    \label{fig:compilation-vs-execution}
\end{figure}

Fig.~\ref{fig:compilation-vs-execution} shows a summary of the compilation and QPU execution time contributions to the total runtime for different optimization levels. Compilation time for the QFT circuit takes between 20\% and 42\% of the total runtime for optimization levels 0, 1, and 2. Interestingly, we observe a decreasing trend in the compilation time for the QFT circuit as we increase the optimization level from 0 to 2, while optimization level 3 increases it up to 47.4 seconds, accounting for 82\% of the total runtime. Notably, the QPU execution time varies between 18.5 and 10.3 seconds, showing a -44\% reduction for increasing the optimization level from 0 to 3.
Conversely, the GHZ circuit compilation time increases with higher optimization levels, accounting for up to 95\% of the total runtime for optimization level 3. In this case, we observed a reduction of the QPU execution time of about -27\% when comparing optimization levels 0 (4.1s) and 3 (3s). However, this marginal tradeoff comes at the expense of substantial compilation time.

\begin{figure*}[!ht]
  \centering
  \begin{subfigure}{0.5\textwidth}
    \includegraphics[width=\linewidth]{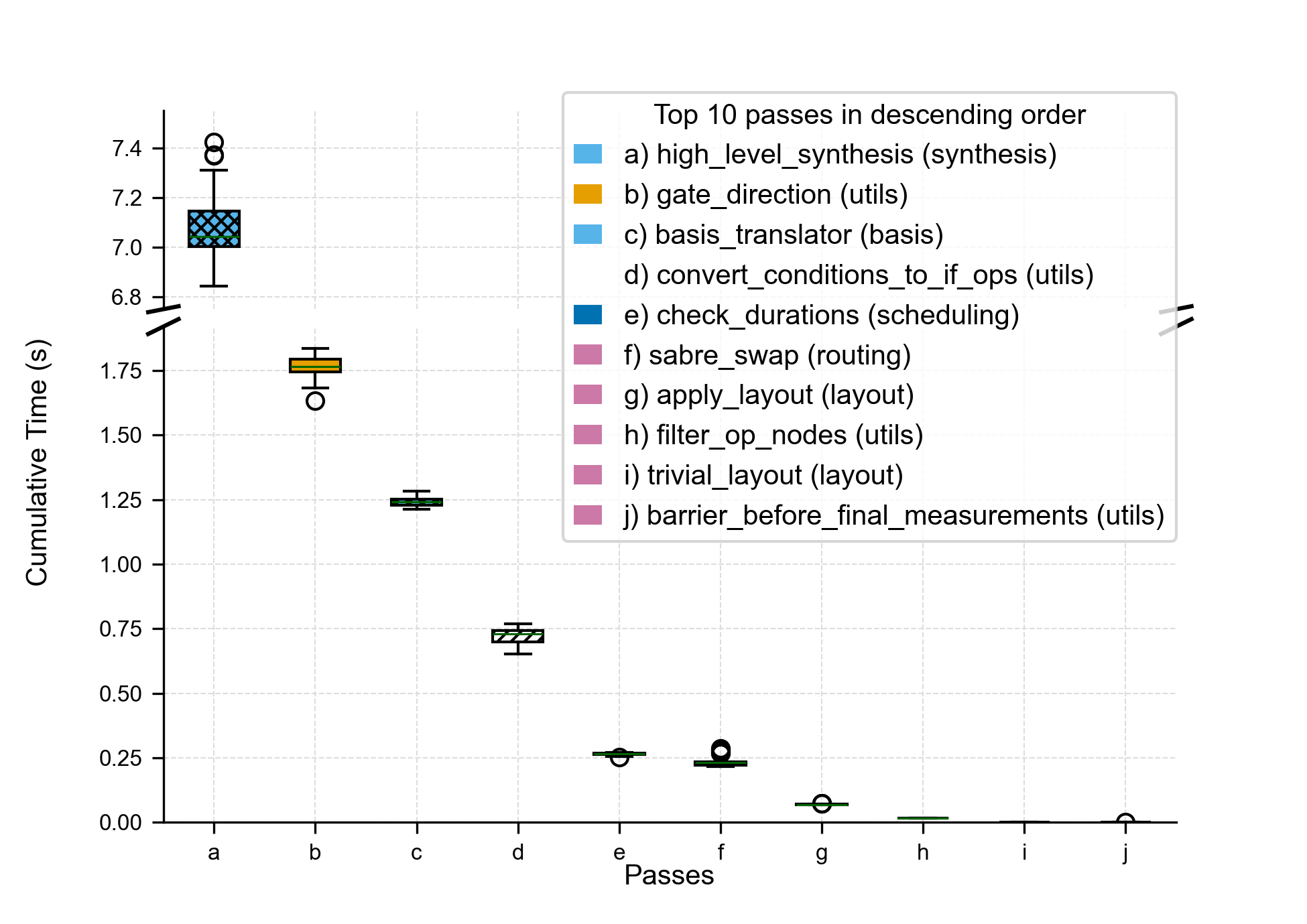}
    \caption{Optimization Level 0}
    \label{fig:sub0QFT}
  \end{subfigure}\hfill
  \begin{subfigure}{0.5\textwidth}
    \includegraphics[width=\linewidth]{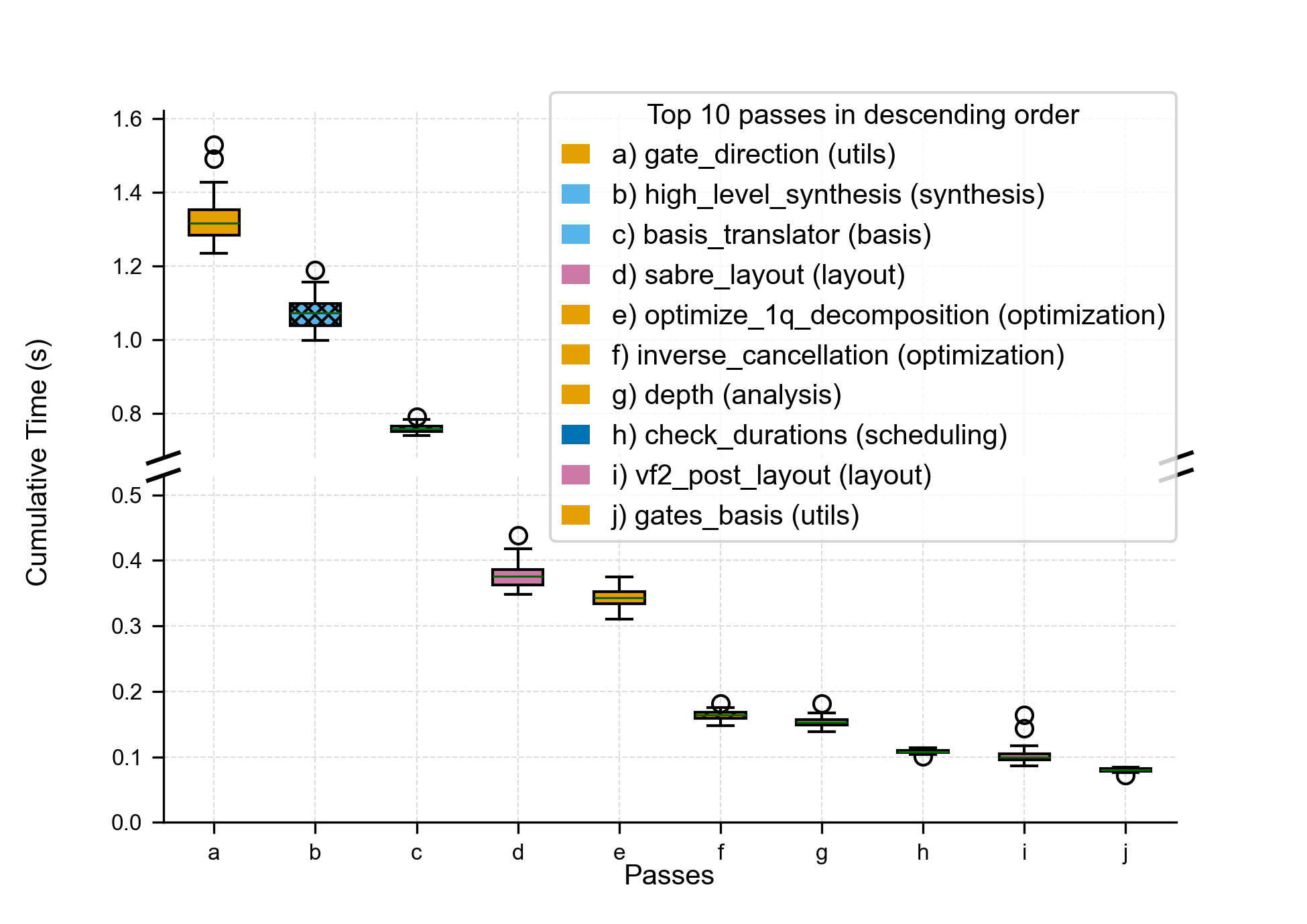}
    \caption{Optimization Level 1}
    \label{fig:sub1QFT}
  \end{subfigure}\hfill

  
  \begin{subfigure}{0.5\textwidth}
    \includegraphics[width=\linewidth]{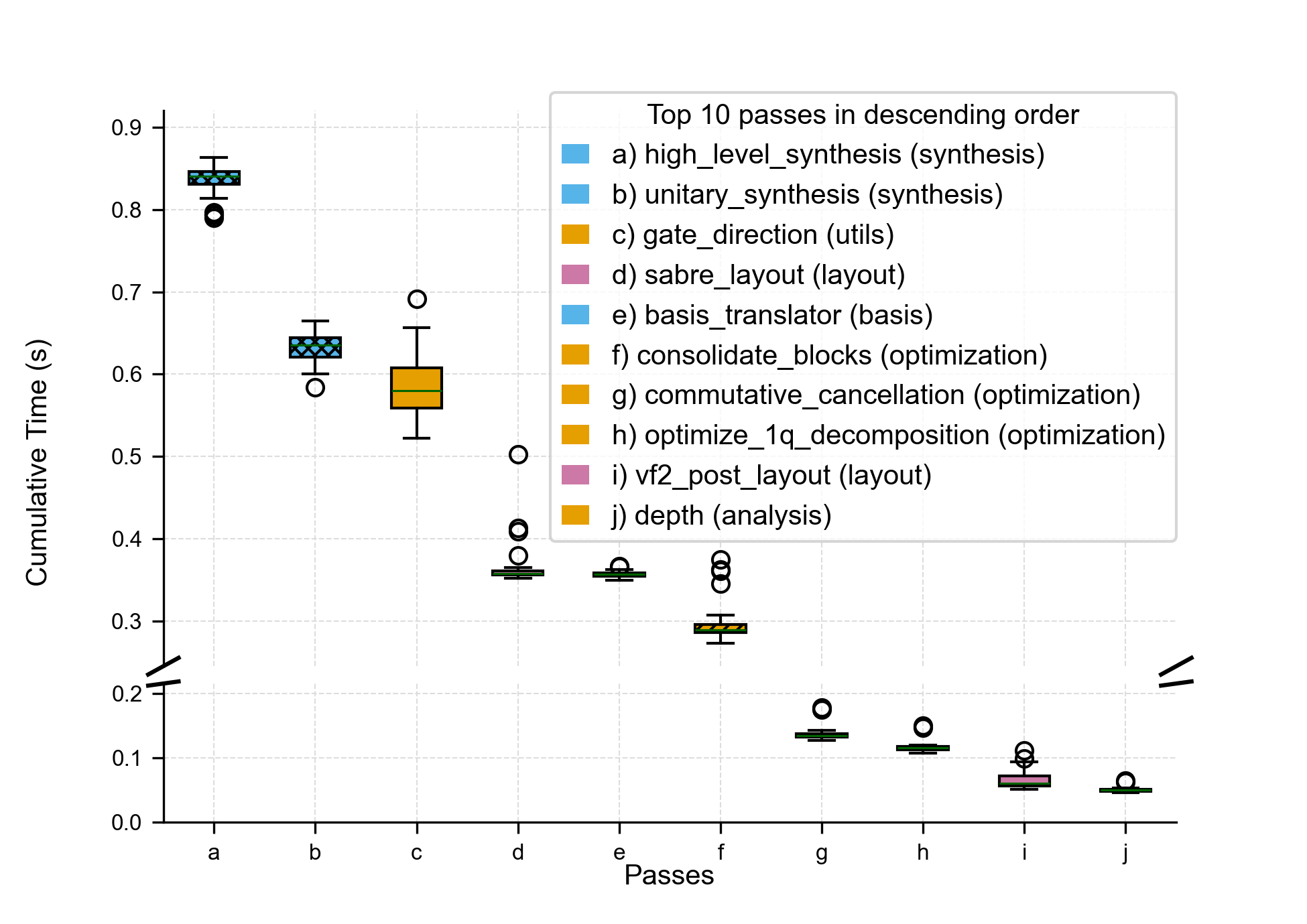}
    \caption{Optimization Level 2}
    \label{fig:sub2QFT}
  \end{subfigure}\hfill
  \begin{subfigure}{0.5\textwidth}
    \includegraphics[width=\linewidth]{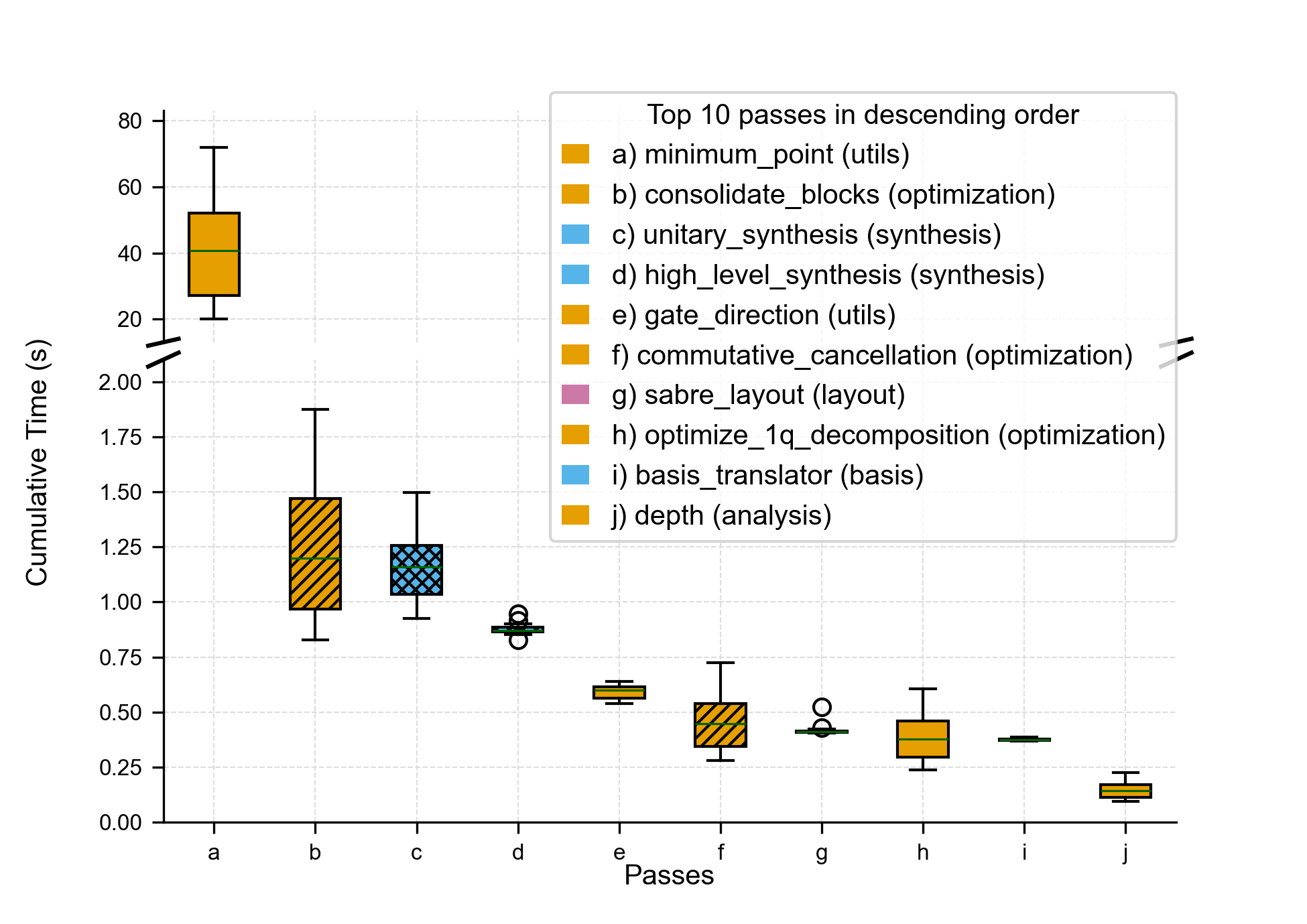}
    \caption{Optimization Level 3}
    \label{fig:sub3QFT}
  \end{subfigure}

  \caption{Boxplots of the top 10 most expensive Qiskit preset compiler passes for the QFT circuit and optimization levels 0 (a), 1 (b), 2 (c), and 3 (d). Passes from the gate synthesis category are marked blue, qubit mapping passes are pink, and circuit optimization passes are orange. Scheduling passes are displayed in dark blue, while uncategorized passes remain white. Right-diagonal hatching (/) indicates passes that occur in two stages, and cross-hatching (\textsf{x}) indicates passes that occur in more than two stages. }
  \label{fig:allplotsQFT}
\end{figure*}

\begin{figure*}[!ht]
  \centering
  \begin{subfigure}{0.5\textwidth}
    \includegraphics[width=\linewidth]{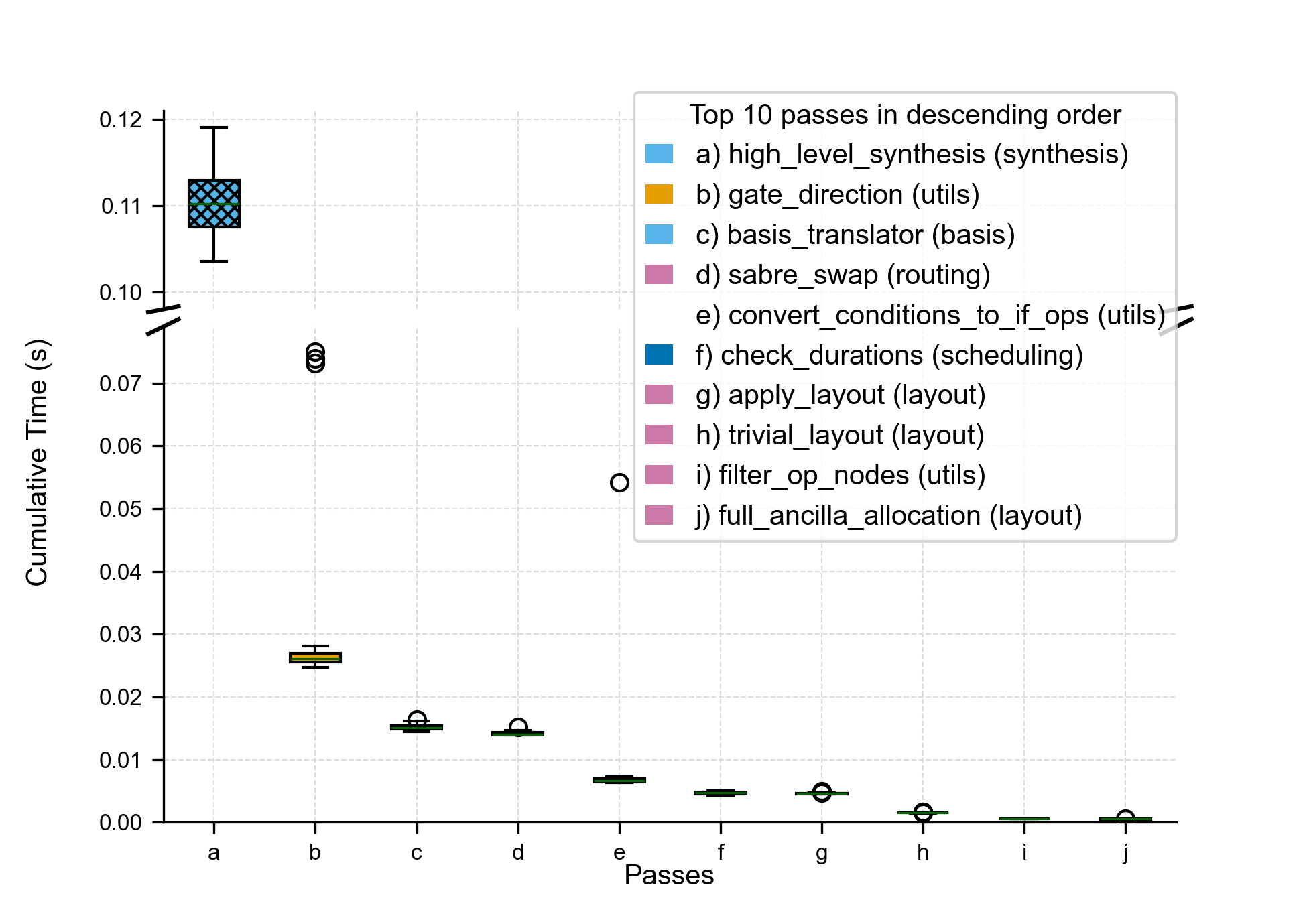}
    \caption{Optimization Level 0}
    \label{fig:sub0GHZ}
  \end{subfigure}\hfill
  \begin{subfigure}{0.5\textwidth}
    \includegraphics[width=\linewidth]{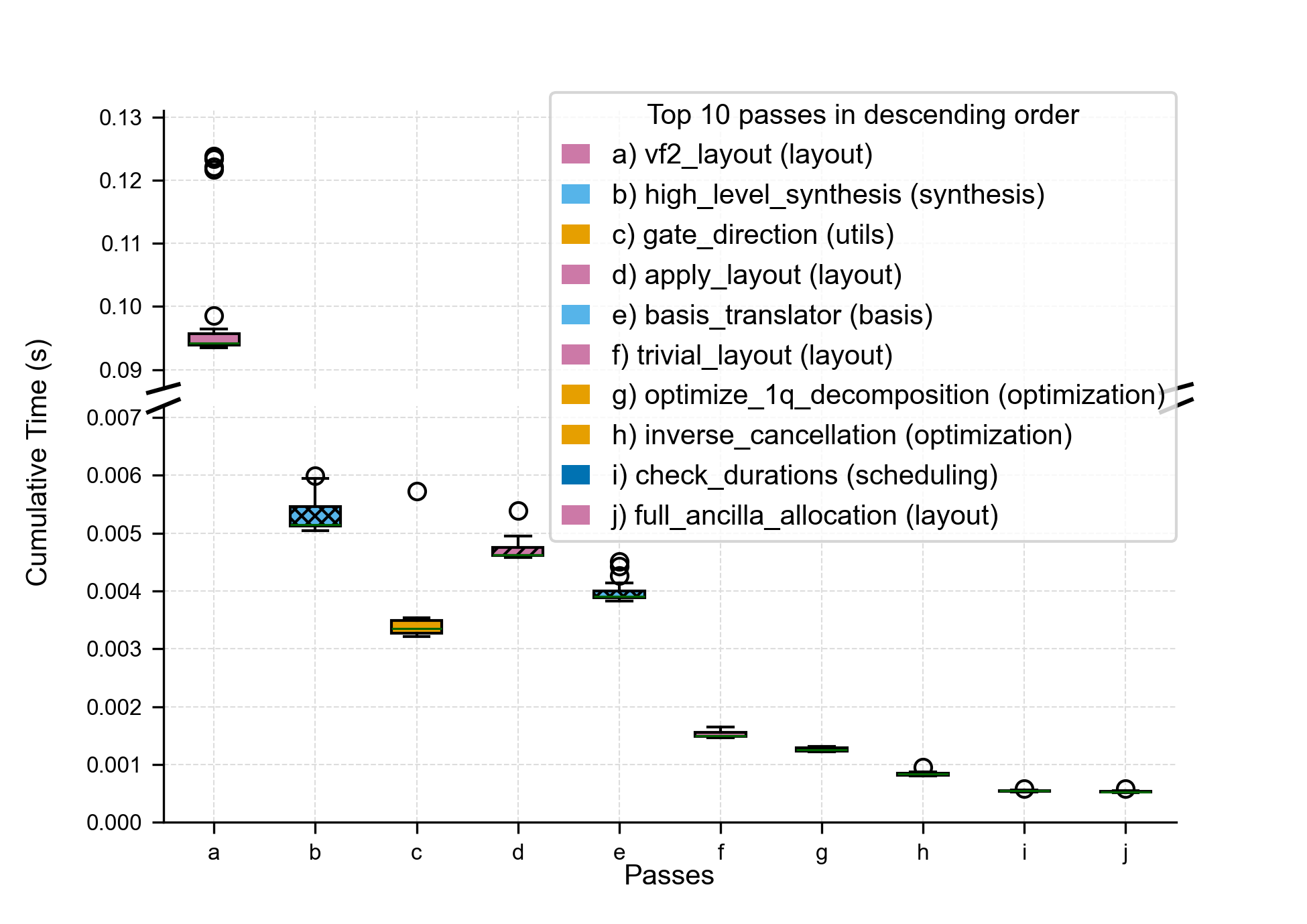}
    \caption{Optimization Level 1}
    \label{fig:sub1GHZ}
  \end{subfigure}\hfill

  
  \begin{subfigure}{0.5\textwidth}
    \includegraphics[width=\linewidth]{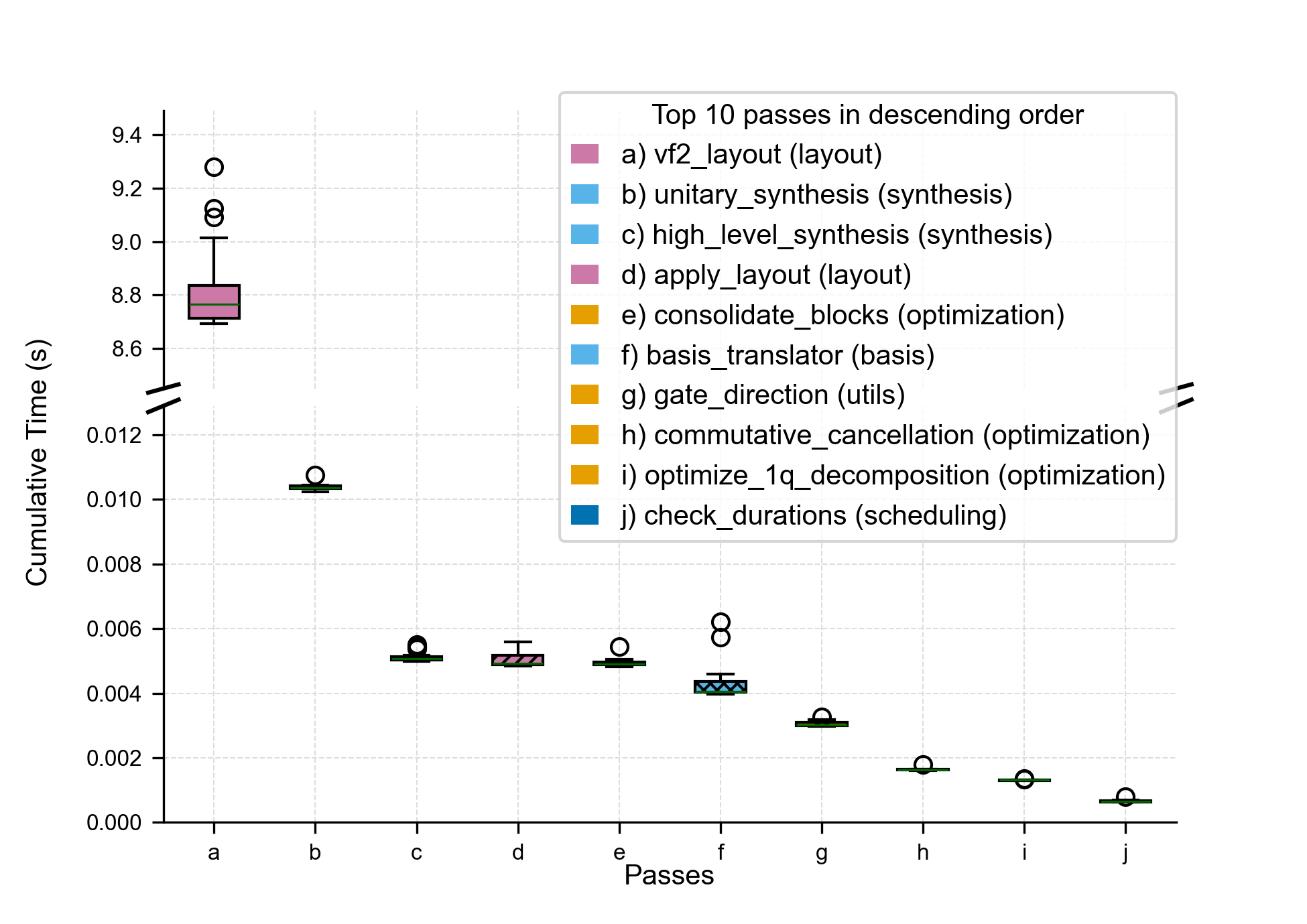}
    \caption{Optimization Level 2}
    \label{fig:sub2GHZ}
  \end{subfigure}\hfill
  \begin{subfigure}{0.5\textwidth}
    \includegraphics[width=\linewidth]{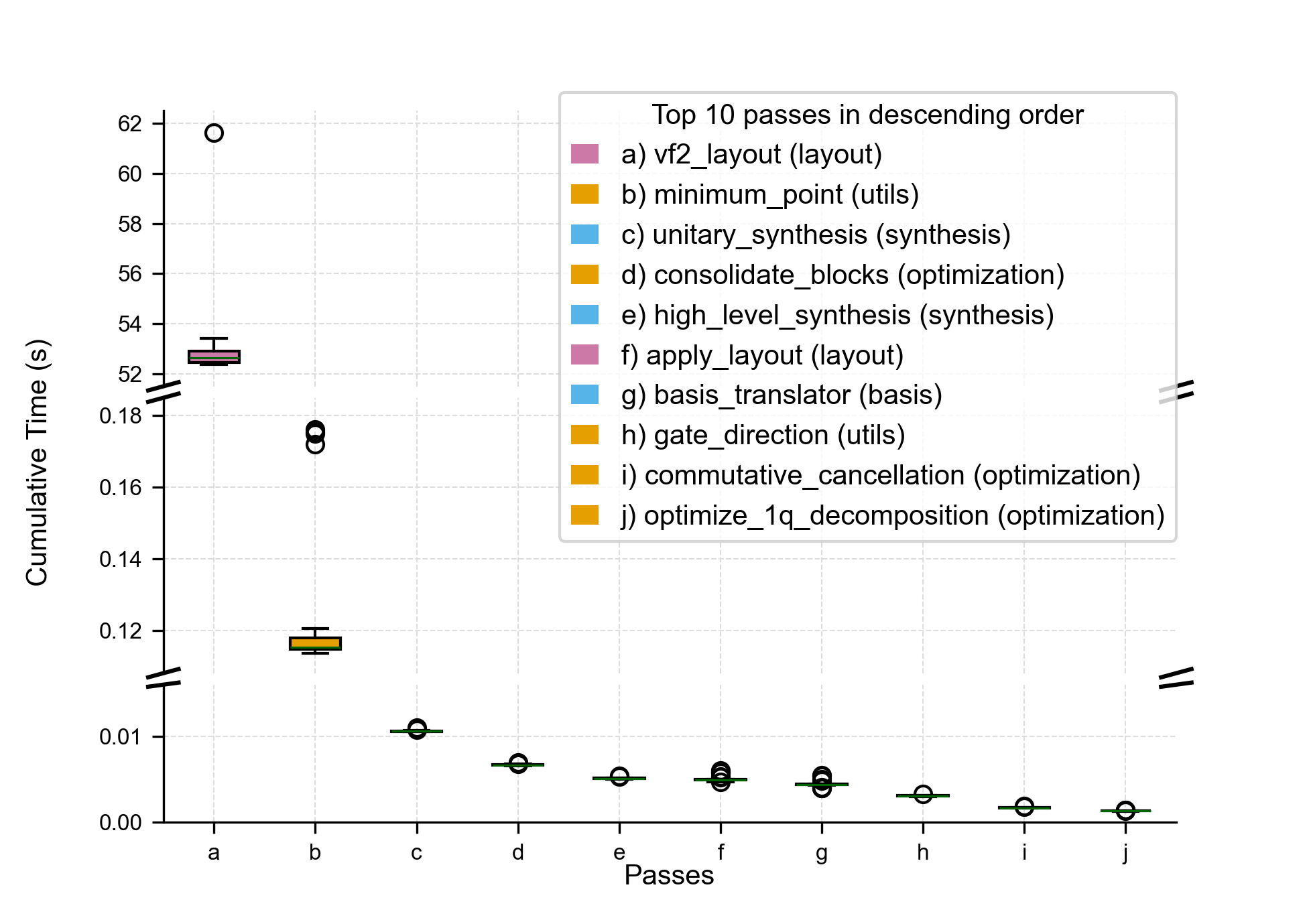}
    \caption{Optimization Level 3}
    \label{fig:sub3GHZ}
  \end{subfigure}

  \caption{Boxplots of the top 10 most expensive Qiskit preset compiler passes for the GHZ state preparation and optimization levels 0 (a), 1 (b), 2 (c), and 3 (d). Passes from the gate synthesis category are marked blue, qubit mapping passes are pink, and circuit optimization passes are orange. Scheduling passes are displayed in dark blue, while uncategorized passes remain white. Right-diagonal hatching (/) indicates passes that occur in two stages, and cross-hatching (\textsf{x}) indicates passes that occur in more than two stages. }
  \label{fig:allplotsGHZ}
\end{figure*}

Figures~\ref{fig:allplotsQFT} and~\ref{fig:allplotsGHZ} show the top 10 most costly compilation passes for the QFT and GHZ circuits, respectively. Both figures are divided into four sub-figures (a)–(d) for each optimization level, and each figure presents box plots of the cumulative time spent executing the corresponding passes. 

The results for the QFT circuit (Fig.~\ref{fig:allplotsQFT}) show that among the top 10 costly passes, five are from the optimization category for optimization levels 1 and 2, and six for optimization level 3. For optimization level 0, a single pass from the same category is observed. Passes from the synthesis category represent two of the top 10 passes for optimization levels 0 and 1, and three of the top 10 for optimization levels 2 and 3. Mapping passes are represented by five out of the top 10 for optimization level 0, but with an increasing optimization level, they are replaced by passes from the optimization category.

The results for the GHZ state preparation circuit (Fig.~\ref{fig:allplotsGHZ}) reveal a comparable trend to that observed in QFT for synthesis passes, with two among the top 10 for optimization levels 0 and 1, and three among the top 10 for optimization levels 2 and 3. The number of optimization passes among the top 10 also increases with respect to the optimization level, as is the case with QFT. Specifically, we observe one, three, four, and five passes for GHZ with optimization levels 0, 1, 2, and 3, respectively. Furthermore, mapping passes consume a substantial portion of the overall runtime in comparison to QFT. In particular, for the GHZ circuit and optimization levels 1, 2, and 3, the \textit{VF2Layout} pass requires several orders of magnitude longer execution times (up to 61.6s) compared to all other passes. With optimization levels 2 and 3, this pass even accounts for over 99\% of total compilation time. 

A notable observation for both circuits is the significantly higher cumulative time for \textit{HighLevelSynthesis} for optimization level 0, which is  $\approx$21x higher in the case of GHZ (up to 0.168s) and $\approx$7x higher in the case of QFT (up to 7.43s) when compared to other optimization levels with the same circuit. Optimization level 0 does not perform high-level optimizations prior to QCC, which is why gate synthesis may require more time. However, it is also important to note that the runtime of synthesis passes is approximately 100x less for GHZ than for the same passes in QFT, which is likely due to the different circuit depths. 

The \textit{MinimumPoint} pass, which occurs only in optimization level 3, consumes a significant amount of time compared to other passes for both circuits, although its impact is stronger with QFT. It is worth noting that for QFT this procedure accounts for about 87\% of total compilation time and can even take up to 72 seconds, the highest cumulative value for an individual pass across all executions. The data used for all figures and results in this work is available via Zenodo\footnote{\href{https://doi.org/10.5281/zenodo.15255700}{https://doi.org/10.5281/zenodo.15255700}}. 


%% file: sections/06-conclusions.tex

\section{Conclusions and Future Work}\label{sec:conclusions}
The primary contribution of this work is a preliminary analysis of the execution profile of the Qiskit built-in compiler toolchain, with the goal of examining the cumulative time spent on each individual pass. Our results represent the first step in our profiling approach to identifying costly passes, which helps researchers and developers detect potential bottlenecks in the compilation process and supports the effective use of precompilation techniques. Our current profiling pipeline measures the cumulative time of a given pass across all stages in which it may occur. As the built-in \texttt{PassManager} from Qiskit executes stages sequentially with no overlap between stages, stage-aware profiling could offer more profound insights into the contributions of each pass per stage. Currently, our analysis is SDK dependent, but similar examinations for other SDKs seem worthwhile. In the future, we aim to extend our analysis to a more extensive set of circuits along with stage-aware profiling and include the actual outcome of the quantum computation in our analysis.